\documentclass[aps,prl,twocolumn,showpacs,superscriptaddress,groupedaddress]{revtex4-1}
\usepackage{amsmath}
\usepackage{amsthm}
\usepackage{amssymb}
\usepackage{mathrsfs}
\usepackage{graphicx}  
\usepackage{dcolumn}   
\usepackage{bm}        
\usepackage{tikz}
\usepackage[margin=1in,footskip=0.25in]{geometry}

\begin{document}

\title{Topological bound states for quantum charges}

\author{I. L. Paiva,$^{1}$ Y. Aharonov,$^{1,2,3}$ J. Tollaksen,$^{1,2}$ and M. Waegell$^{2}$}
\affiliation{$^{1}$Schmid College of Science and Technology, Chapman University, Orange, California 92866, USA}
\affiliation{$^{2}$Institute for Quantum Studies, Chapman University, Orange, California 92866, USA}
\affiliation{$^{3}$School of Physics and Astronomy, Tel Aviv University, Tel Aviv 6997801, Israel}

\begin{abstract}
We discuss how, in appropriately designed configurations, solenoids carrying a semifluxon can be used as topological energy barriers for charged quantum systems. We interpret this phenomenon as a consequence of the fact that such solenoids induce nodal lines in the wave function describing the charge, which on itself is a consequence of the Aharonov-Bohm effect. Moreover, we present a thought experiment with a cavity where two solenoids are sufficient to create bound states.
\end{abstract}

\maketitle

The Aharonov-Bohm (AB) effect reveals a surprising influence of topology on the dynamics of quantum systems \cite{Aharonov1959}. In its most known form, it shows that the evolution of a quantum system around an infinite solenoid is affected by the magnetic flux on it, even thought the solenoid only produces an electromagnetic field on its interior. More specifically, if a quantum wave packet with charge $q$ travels around a solenoid carrying a magnetic flux $\Phi_B$ in its interior, it accumulates a geometric phase whose modular part corresponds to
\begin{equation}
\Delta\phi = \frac{q\Phi_B}{\hbar c}. \label{AB effect}
\end{equation}
Experimental validations of this effect and its consequences and interpretation have been discussed extensively in the literature -- see, e.g., Refs. \citep{Peshkin1961, Liebowitz1965, Boyer1973, berry1980, Olariu1985, tonomura1986evidence, osakabe1986experimental, berry1989quantum, Peshkin1989, aharonov1994, recher2007aharonov, russo2008observation, Aharonov2009, peng2010aharonov, berry2010semifluxon, Kaufherr2011, Vaidman2012, fang2012photonic, bardarson2013quantum, noguchi2014aharonov, duca2015aharonov, mukherjee2018experimental,cohen2019geometric}. Most of the applications make straightforward use of the AB effect, with particles encircling a region with magnetic flux \cite{recher2007aharonov, russo2008observation, peng2010aharonov, bardarson2013quantum, noguchi2014aharonov, duca2015aharonov, mukherjee2018experimental}. Here, however, we investigate how the AB effect changes stationary energy eigenstates by altering the topology of the space of solutions to the Schr\"odinger equation, following the methods developed in \citep{aharonov1994}. We show that solenoids carrying half-integer fluxons, or {\it semifluxons} \citep{berry2010semifluxon}, can be used to confine a charge within a certain spatial region. Moreover, this effect can be used to confine a charge between two such solenoids in a small sector of a long cavity. We call the states of those confined particles {\it topological bound states} \footnote{We call these topological bound states because the confinement effect results from the way the AB effect changes the topology of the solution space of the Schr{\"o}dinger equation. We should stress that, in principle, this has no direct connection to uses of the word topology in other areas of physics.}.


Using scattering theory, it has been noted in previous studies of single-channel transport that the presence of an AB ring with a semifluxon can lead to zero transmission \citep{ryu1998phase, lee1999generic, taniguchi1999friedel, kim2002effects}.  Here, our discussion provides an explanation of this effect in terms of topology of solutions of the Schrodinger equation, geometry, and energetics.


In this letter, as is typically done, solenoids are represented by single points on planes and only two-dimensional versions of the settings are considered. To set our framework, we consider a charged particle inside a cavity of arbitrary shape which contains a solenoid with no magnetic flux at a certain point $P$. If the particle's wave function $\psi_0$ is an eigenstate of this cavity with energy $E$, then it satisfies the Schr\"{o}dinger equation
\begin{equation}
-\frac{\hbar^2}{2m}\nabla^2\psi_0=E\psi_0,
\label{eq-simple}
\end{equation}
where $m$ is the mass of the charge. In this case, we assume $\psi_0$ is real without loss of generality. Now, let the solenoid inside the cavity carry an arbitrary magnetic flux. If $\psi$ is an eigenstate of energy, it satisfies
\begin{equation}
-\frac{1}{2m}\left(\hbar\nabla-i\frac{q}{c}A\right)^2\psi=E\psi,
\label{eq-pot}
\end{equation}
where $A$ is the vector potential associated with the solenoid. One can easily verify that
\begin{equation}
\psi = \psi_0 e^{i\frac{q}{\hbar c}\int_\gamma A \cdot d\gamma},
\label{psi}
\end{equation}
where $\gamma$ denotes a possible path. Equation \eqref{psi} can, then, be seen as a complex extension rule for $\psi_0$, which then determines a Riemann surface \cite{forster2012lectures}. In other words, $\psi$ is a ``multivalued function'' with evaluation at a certain point $r$ given by
\begin{equation}
\psi(r) = \psi_0(r)e^{iW_\gamma\Delta\phi},
\label{ext-rule}
\end{equation}
where $W_\gamma$ is the winding number of $\gamma$, since \cite{berry1986statistics}
\begin{equation}
\oint_\gamma A \cdot d\gamma = W_\gamma \Phi_B.
\end{equation}
In order to obtain a single-valued function, a sheet of such a manifold has to be chosen. However, this choice introduces an arbitrary discontinuity in the wave function. Nonetheless, we note that this discontinuity is only in the system's phase. In fact, this approach of making $\psi$ single-valued can be intuitively seen as an imposition of a periodic boundary condition on $\psi_0$. Then, instead of solving \eqref{eq-pot}  for $\psi$, one solves \eqref{eq-simple} for $\psi_0$ with the extra condition that, in polar coordinates $(\rho,\theta)$,
\begin{equation}
\psi_0(\rho,\theta+2\pi) = e^{-i\Delta\phi} \psi_0(\rho,\theta).
\end{equation}
This is physically relevant because it implies that the absolute value of the wave function and also its probability density are independent of the chosen sheet of the manifold, as expected.

As already mentioned, the solenoids we consider carry a {\it semifluxon}, i.e., a magnetic flux given by
\begin{equation}
\Phi_B = \frac{\pi\hbar c}{q}.
\label{semifluxon}
\end{equation}
This choice is made because it makes the extension rule \eqref{ext-rule} real, since $\Delta\phi=\pi$ and $e^{i\Delta\phi}=-1$, which allows us to take $\psi$ real. For any other value of $\Phi_B$, the wave function $\Psi$ is manifestly complex, and the reasoning we use in this work does not follow. For that reason, we cannot say that our results are robust against fluctuations in the magnetic flux carried by the solenoid. Also, because $\psi$ gets a phase of $\pi$ in a path which encircles a semifluxon, it must be zero somewhere in the path, regardless of the shape the cavity. Therefore, by the continuity of the surface determined by $\psi$ (or the continuity of its probability distribution in case a sheet is chosen), there must be a nodal curve (i.e., a curve where $\psi$ vanishes) starting at the solenoid and finishing at the boundary of the cavity (or at the position of another solenoid carrying a semifluxon) \cite{aharonov1994, berry2010semifluxon}. Nodal curves appear in a variety of topological systems, not only in systems with solenoids. Recently, special attention has been dedicated to them and they have been realized in multiple physical configurations \cite{kim2015dirac, bian2016topological, bzduvsek2016nodal, hirayama2017topological, takahashi2017spinless, gao2018experimental, yan2018experimental, xia2019observation}. We also note that these nodal curves of eigenstates become straight lines in particularly designed symmetric geometries. For the results considered in this work, such symmetries are always present. Moreover, we note that linear combinations of eigenstate also possess nodal curves, which can be seen by taking linear combinations of different Riemann surfaces.

\begin{figure}[!ht]
\begin{center}
\includegraphics[scale=.11]{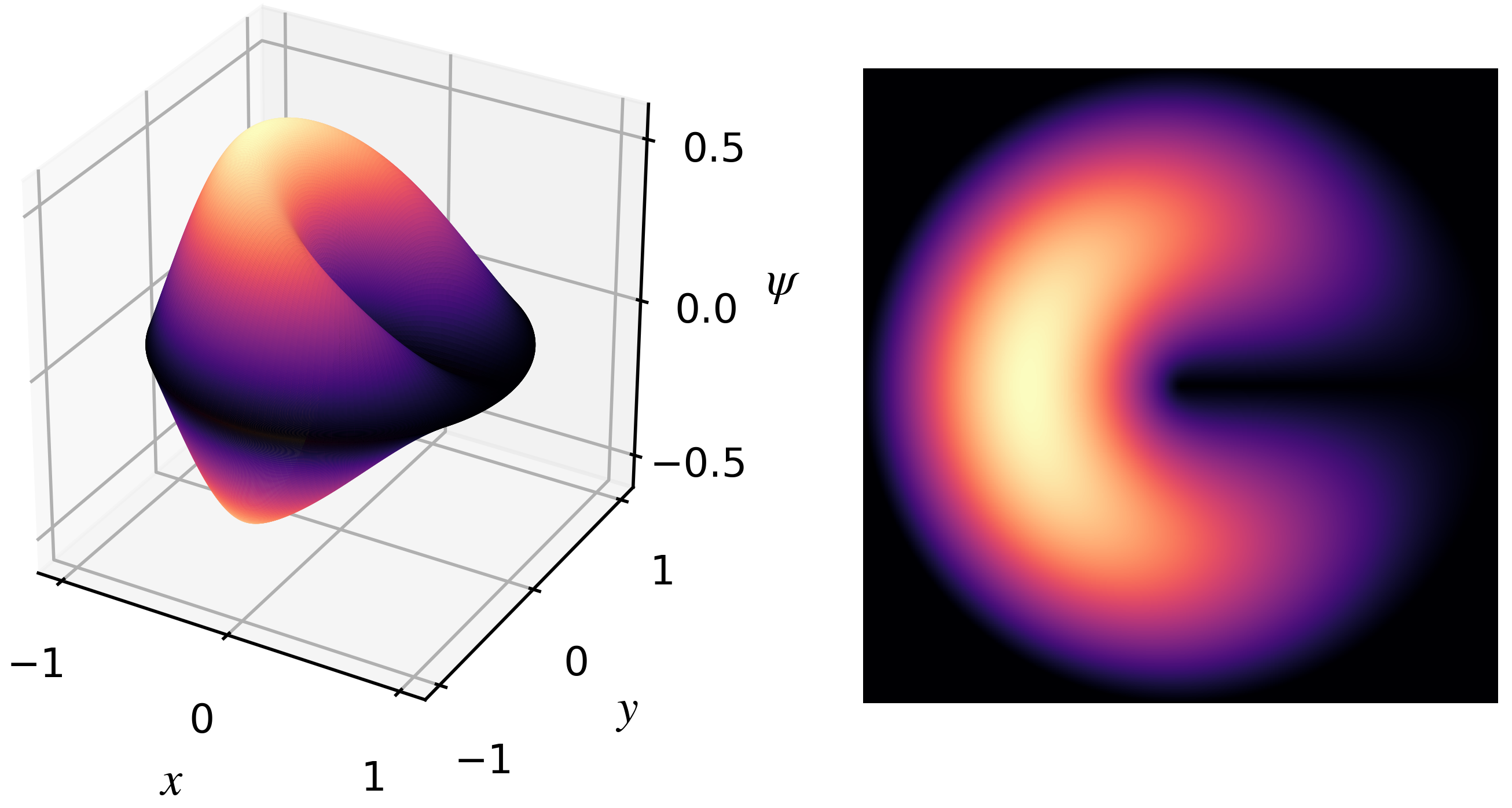}
\end{center}
\caption{Representation of the real ground state of a circular cavity with a solenoid carrying a semifluxon on its center. The three-dimensional plot on the left is the real Riemann surface associated with the ``multivalued'' wave function of Eq. \eqref{ext-rule} for $\Delta\phi=\pi$. The color map defined on it associates colors to the magnitudes of $\psi$. This allows us to draw the heat map of any arbitrary chosen sheet of that surface. Such a choice is used to make $\psi$ a single-valued function.}
\label{fig0}
\end{figure}

To summarize and illustrate the main points of our discussion up to now, we consider a circular cavity of unit radius with a solenoid carrying a semifluxon at its center. In this case, the ground state of the cavity when $A=0$ is $\psi_0(r)=J_1(\alpha \rho)$, where $\rho$ is the radial coordinate, $J_1$ is the Bessel function of first kind and parameter 1, and $\alpha\approx3.83$ is the smallest positive zero of $J_1$. Then, when the solenoid is carrying a semifluxon, the real ground state can be expressed as $\psi(r)=\psi_0(r)\cos(\theta/2)$, where $\theta$ is the angular coordinate. Note that $\psi$ is a ``multivalued function''. On the left-hand side of Fig. \ref{fig0}, the Riemann surface associated with $\psi$ is represented, while the heat map of any arbitrary chosen sheet of the surface is represented on the right-hand side. Observe the nodal line induced by the solenoid.

Now, if the charge is in free space, the analysis we made so far is still valid. In this case, a nodal line starts at a solenoid and finishes at infinity -- or at the location of another solenoid carrying a semifluxon.

For our main result, we associate to these nodal lines the fact that semifluxons behave as topological energy barriers for charged particles. To see that, we revisit the scattering experiment of a charge incident on a grating of slits and a lattice of solenoids \citep{Aharonov1959,Aharonov2005}. First, consider a particle with charge $q$. Moreover, assume it is moving in the horizontal direction ($x$-axis) with momentum $p=2\pi\hbar/\lambda$. At $x=0$, the system encounters a vertical ($y$-axis) infinite grating consisting of narrow slits with a distance $L$ between them. We approximate the incoming packet as a plane wave in y.

If the initial state of the charge, which has zero transverse ($y$-axis) momentum, is represented by $|\psi_i\rangle$, and $|\psi_f\rangle$ is its final state -- with transverse momentum given by $\hbar k_f$ -- the first order perturbation of the transition probability is proportional to
\begin{equation}
\langle \psi_f|V(y)|\psi_i\rangle \propto \int_{-\infty}^{\infty} e^{-i k_f y} V(y) dy,
\end{equation}
where $V(y)$ is the disturbing potential that represents the grating of slits and hence is periodic in $y$ with period $L$. Because of this, the probability amplitude does not vanish if and only if
\begin{equation}
k_f=\frac{2\pi n}{L}, \qquad n\in \mathbb{Z}.
\end{equation}

The above result shows that there is a {\it quantized} exchange of momentum between the grating and the charge. Moreover, the particle gets a quantized momentum in the $y$-direction given by
\begin{equation}
p_y = \hbar k_f = \frac{2\pi n\hbar}{L}. \label{py}
\end{equation}

If we assume the grating is heavy, then its average energy should remain unchanged. Therefore, the average energy of the particle is unaltered by its interaction with the slits, which, in turn, leads to the conservation of the magnitude of its average momentum. This idea together with expression (\ref{py}) leads to the well-known scattering angle
\begin{equation}
\sin\theta_n = \frac{p_y}{p} = \frac{n\lambda}{L}. \label{scatter1}
\end{equation}

\begin{figure}[!ht]
\begin{center}
\includegraphics[width=\columnwidth]{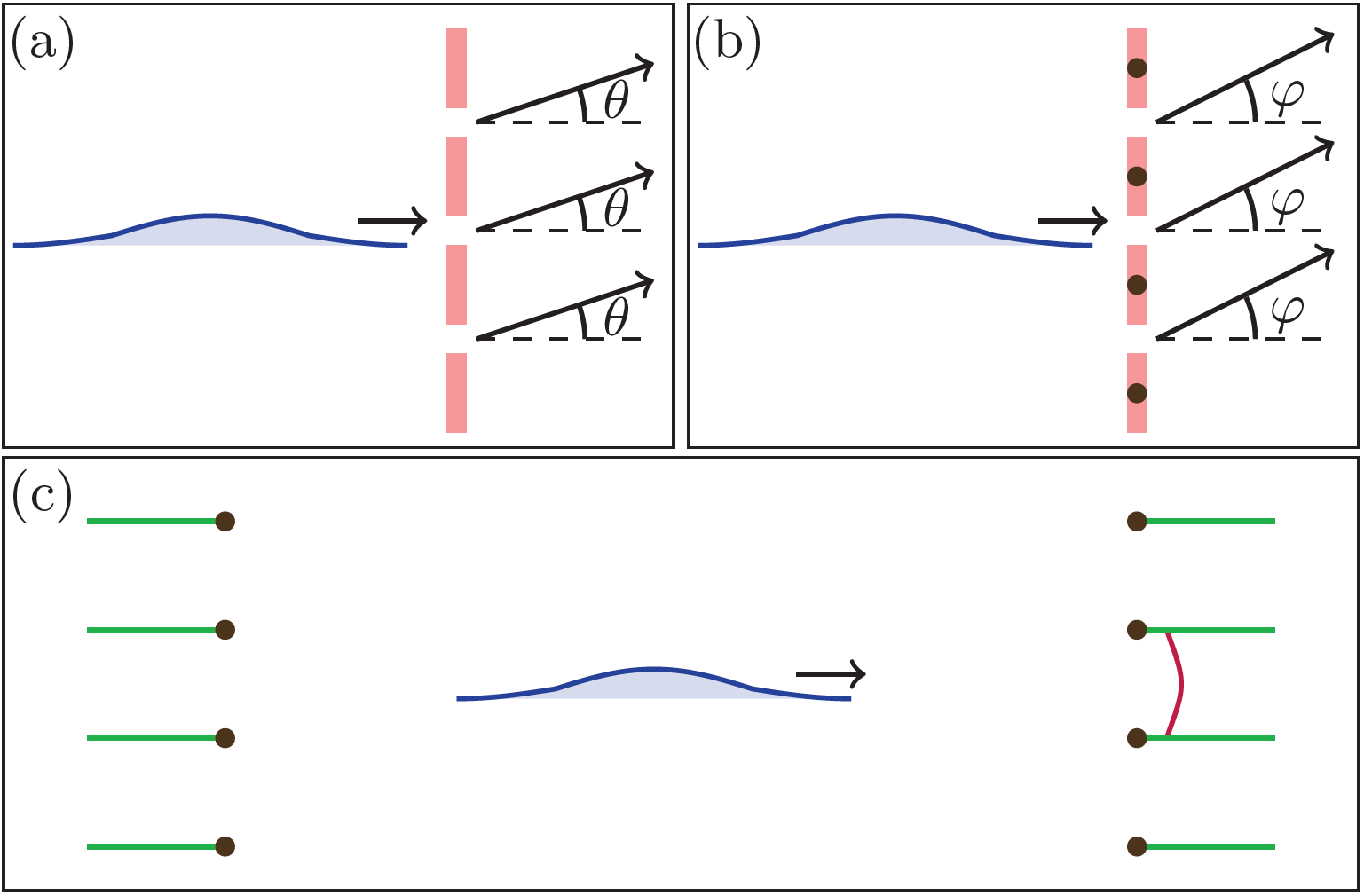}
\end{center}
\caption{Schematic representation of scattering experiments. The dark dots are solenoids carrying semifluxons. The scattering angle changes with the addition of the lattice of solenoids. Moreover, when a charge is incident on the apparatus, part of the wave function is reflected back in (b) and gets trapped in the region between the two lattices of solenoids in (c). The green lines in (c) represent the nodal lines induced by the solenoids and the red curve ``escaping'' the region between the lattice of solenoids represent the state with the smallest energy in the $y$-direction that can do so.}
\label{fig1}
\end{figure}

Note that, even though there is an exchange of momentum in the $y$-direction, one does not expect a change in the average momentum of the charge -- nor of the grating. As an aside, interactions of this type, which do not change the mean value of observables, can be examined using the concept of {\it modular variables} \citep{Aharonov1969, Aharonov2005, popescu2010dynamical, Lobo2014, Aharonov2017}.

Now, consider a modified version of this experiment where a semifluxon is placed between each neighboring pair of slits, forming a lattice of solenoids, as shown in Fig \ref{fig1}(b). We assume the lattice of solenoids moves as a single rigid object. Also, let the grating of slits be free to move independently of the lattice of solenoids.

In this scenario, the scattering angle gets shifted by the phase $\Delta\phi=\pi$ due to the AB effect. In fact, the new angle is given by \citep{Aharonov1959}
\begin{equation}
\sin\varphi_n = \left(n+\frac{1}{2}\right)\frac{\lambda}{L}.
\end{equation}

This shift in the scattering pattern is the crux of our result. It implies that the amount of momentum the particle gets in the $y$-direction is
\begin{equation}
p_y=p\sin\varphi_n=\left(n+\frac{1}{2}\right)\frac{2\pi\hbar}{L}.
\end{equation}
Hence, the particle gets a transverse momentum with magnitude of at least $\pi\hbar/L$. Here, again, note that the average transverse momentum should be unchanged.

As a result, the solenoids introduce an energy barrier to the charge. To make this clear, note that if the initial momentum $p$ of the charge in the $x$-direction is much smaller in magnitude than $\pi\hbar/L$, the system can only go through the lattice via tunneling in a way that conserves its average energy. Hence, there is a rapid decay in the probability to find the charge beyond the lattice. That is, the probability of finding the particle decreases the further a screen is put to detect it on the right-hand side of the lattice of solenoids. Then, in general, measurements of the final position of the system in repeated experiments would show that the charge bounced back off the lattice most of the time.

Furthermore, because the grating of slits and the lattice of solenoids are independent, we can modify the experiment by removing the grating of slits. The energy barrier remains.

Next, we show a way to obtain this result, which makes use of the fact that the solenoids carrying semifluxons ``force'' nodal lines in the eigenstates. First, we point out that given the symmetry of the problem and the initial state of the charge, the nodal line associated with each solenoid must be horizontal. This holds because the particle is moving in the $x$-direction and the lattice of solenoids stays invariant by translations of length $L$ in the $y$-direction. Moreover, if we assume that the particle starts localized on the left-hand side of the lattice without a nodal line and moving as pulse with negligible spreading, the decomposition of the initial state of the particle into eigenstates of the Hamiltonian should have most of its contribution from eigenstates with nodal lines that start on the right-hand side of each solenoid and go to infinity, as represented by the green lines on the right-hand side of Fig. \ref{fig1}(c). Then, the portion of the initial wave function associated with average energy greater than the energy of the ground state of an infinite square well with length $L$ -- or average momentum $\left|\langle p\rangle\right|\geq \pi\hbar/L$ -- is transmitted through the lattice. The remaining part is reflected back.

Consider now a similar lattice of solenoids placed on a vertical line which is far away from the original lattice on the left-hand side of it, as represented by Fig. \ref{fig1}(c). In this setting, the parts of the initial wave packet that were reflect off the first lattice of solenoids are likewise reflected by the new lattice. The result is a topologically trapped charge, which is in a superposition of {\it topological bound eigenstates}. We emphasize again that topological bound eigenstates have tails bounded by an exponential decay outside the region between the solenoids due to the continuity of the probability density associated with the wave function.

Now, we present our final result, which provides a construction of topological bound states that might actually be implemented in the the laboratory. Consider a rectangular cavity equipped with two solenoids, each carrying a semifluxon, placed with a distance $L$ from each other on the long symmetry axis of the cavity. Let the height of the cavity be $d$ and its width be $D$, with $D \rightarrow \infty$ corresponding to an open wave-guide. We are interested in the case $D \gg L \gg d$, as explained next.

\begin{figure}[!ht]
\includegraphics[width=\columnwidth]{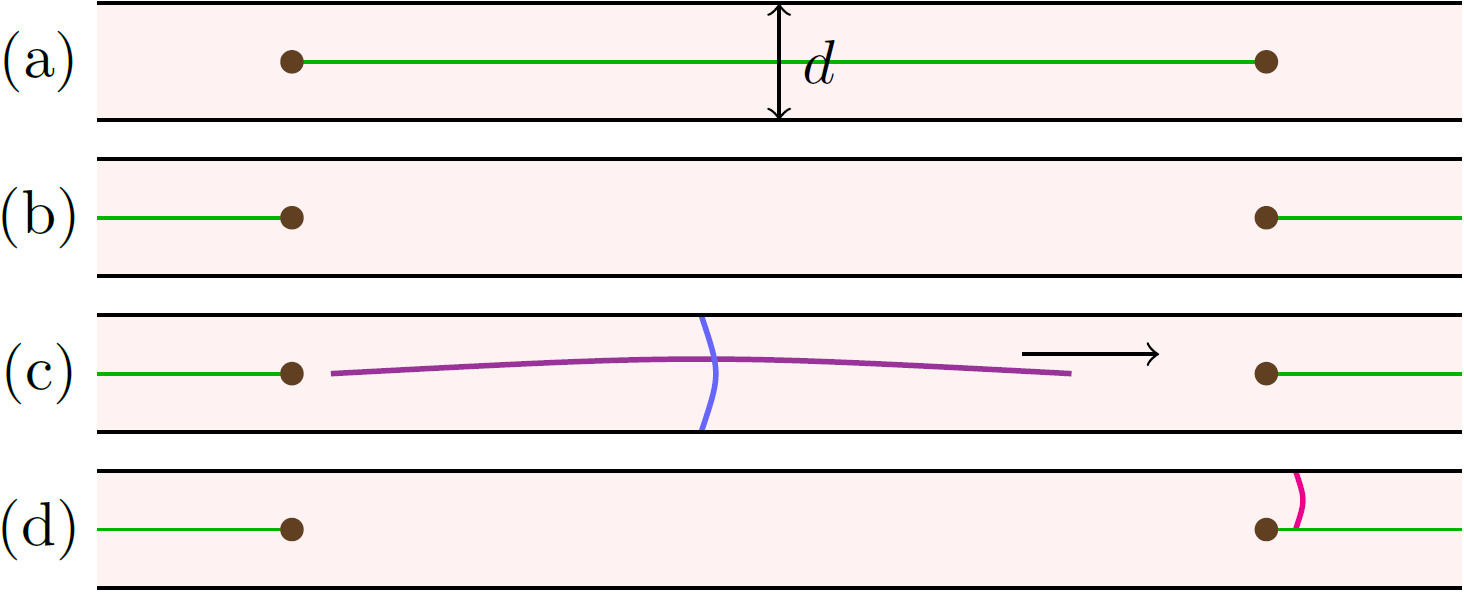}
\caption{Long cavity with two solenoids (dark dots), each carrying a semifluxon. The green lines correspond to nodal lines induced by the solenoids. In (c), a product state of the ground state of the cavity without solenoids in the vertical and a state with low average energy in the horizontal is represented. Part of this system must stay confined in the sector delimited by the solenoids because a system that completely moves past that sector must have at least the energy of the first excited y-mode of an identical cavity without the solenoids, as represented by the red curve in (d).}
\label{cavity}
\end{figure}

If the charge is initially symmetrically distributed along the vertical direction, the symmetry of the problem implies that the nodal lines associated with the solenoids for eigenstates of the cavity should be as illustrated by the green lines in either Figs. \ref{cavity}(a) or \ref{cavity}(b). To see that, note that the Hamiltonian of the cavity remains physically invariant if the cavity is rotated by $\pi$ about its long axis of symmetry. In fact, the difference between the two situations is the direction of the semifluxon. Then, the flux on the solenoid differs by the value of a fluxon, which does not affect any physical system. Eigenstates with low energy associated with the configuration presented in Fig. \ref{cavity}(b) are mostly confined in the region between the solenoids and any superposition of these topological bound eigenstates is a topologically trapped state, which we now show in more detail.

Assume the initial wavefunction is a packet confined in the region between the two solenoids and symmetrically distributed along the vertical direction, with momentum towards the right, and negligible spreading. Moreover, assume its average energy $\langle E\rangle$ is much smaller than the first excited state of the infinite square well with length $d$, i.e.,
\begin{equation}
\langle E\rangle \ll \frac{2\pi^2\hbar^2}{md^2}.
\label{ground energy}
\end{equation}
This situation is represented in Fig. \ref{cavity}(c). Because the distance between the solenoids can be taken to be sufficiently large, it is possible, by the uncertainty principle, to construct such a state with low uncertainty in momentum and, hence, in energy.

In this setup, the eigenstates that contribute the most to the eigendecomposition of the particle must be the ones whose nodal lines are in the configuration shown in Fig. \ref{cavity}(b). This, in turn, ``forces'' the average energy of a state that completely escapes the region between the solenoids to be at least the energy of the first excited state of the infinite square well with length $d$, as seen in Fig. \ref{cavity}(d).

As a consequence, even if parts of the original state of the charge associated with higher energy move past the solenoid on the right-hand side, parts of it bounce back. This remaining part, once again, bounces back after it attempts to go past the solenoid on the left-hand side. Hence, this part of the state is trapped in the region between the solenoids. It is, therefore, a topologically trapped state given by a linear combination of topological bound eigenstates with nodal lines outside the region between the solenoids.

In the cavity, as in the case of the lattice of solenoids, the continuity of the probability density associated with the wave function implies that the topological bound states have exponentially decaying tails outside the region between the solenoids. Therefore, position measurements of topologically bounded charges must reveal with large probability that they are in the region between the solenoids -- even if $D>\infty$. To decrease the odds of finding the charge outside this region, the distance between the solenoids should be increased.

We reiterate that the inclusion of the two semifluxon solenoids as described creates energy barriers inside the cavity. These energetic barriers extend to the boundary of the cavity, which makes this situation analogous to the case of a finite potential well, where a constant finite potential is added everywhere outside a certain region, and the topological bound states analogous to the bound states of the finite square well. As a matter of fact, finite square wells could also be used to trap quantum systems with sufficiently low energy in a certain subregion of a cavity. However, our trapping method might be easier to implement, since it only requires changing the current in the two solenoids.


To summarize, in this article we have shown that the AB effect enables the construction of topological barriers that can be used to trap charged particles. In particular, a quantum state within a cavity can be further confined to specific sectors of it. This is a non-trivial use of the AB effect since we do not consider systems dynamically encircling a region with magnetic field. Such cavities with solenoids can be implemented experimentally, for instance, by using synthetic magnetic fields, where the solenoid fields can be created optically. Moreover, in future works, generalizations for other values of magnetic flux inside the solenoid can be investigated. Furthermore, other configurations could be considered in a search to confine charges with a finite number of solenoids without the use of auxiliary systems such as the cavity we considered here. Finally, although the ideas discussed here are only valid for charges, one could look for similar results for neutral systems with magnetic moments since they can also be subject to the AB effect \citep{Aharonov1984}.

\section{Acknowledgments}
We thank the referees for their careful reading of our manuscript and for their constructive comments, in particular, for the suggestion of Ref. \citep{kim2002effects}, which made us aware of works on transmission zeros in channels with AB rings. This research was partially supported by the Fetzer Franklin Fund of the John E. Fetzer Memorial Trust. I.L.P. acknowledges financial support from the Science without Borders Program (CNPq, Brazil). Y.A. acknowledges support from the Israel Science Foundation (Grant 1311/14), Israeli Centers of Research Excellence (ICORE) Center ``Circle of Light,'' and the German-Israeli Project Cooperation (Deutsch-Israelische Projektkoopera-tion, DIP).

\bibliography{paper}
\bibliographystyle{apsrev4-1}
\end{document}